# A Labelled Dataset for Sentiment Analysis of Videos on YouTube, TikTok, and Other Sources about the 2024 Outbreak of Measles

Nirmalya Thakur[1], Vanessa Su[2#], Mingchen Shao[1#], Kesha A. Patel[2†], Hongseok Jeong[1†], Victoria Knieling[3†], and Andrew Bian[4†]

[1] Department of Computer Science, Emory University, Atlanta, GA 30322, USA
[2] Department of Mathematics, Emory University, Atlanta, GA 30322, USA
[3] Program in Linguistics, Emory University, Atlanta, GA 30322, USA
[4] Goizueta Business School, Emory University, Atlanta, GA 30322
nirmalya.thakur@emory.edu
vanessa.su@emory.edu
katie.shao@emory.edu
kesha.patel@emory.edu
peter.jeong@emory.edu
victoria.knieling@emory.edu
andrew.bian@emory.edu

**Abstract.** Since the beginning of 2024, several countries have been experiencing an outbreak of measles. In the modern-day Internet of Everything lifestyle, social media platforms such as YouTube and TikTok have gained widespread popularity on a global scale due to their ability to facilitate the easy creation and dissemination of videos. During virus outbreaks of the recent past, videos on social media platforms played a crucial role in keeping the global population informed and updated regarding various aspects of the outbreaks. As a result in the last few years, researchers from different disciplines have focused on the development of datasets of videos published on YouTube, TikTok, and similar websites. No prior work in this field has focused on the development of a dataset of videos about the ongoing outbreak of measles, published on social media platforms. The work of this paper aims to address this research gap and presents a dataset that contains the data of 4011 videos about the ongoing outbreak of measles published on 264 websites on the internet between January 1, 2024, and May 31, 2024, available at https://dx.doi.org/10.21227/40s8-xf63. These websites primarily include YouTube and TikTok, which account for 48.6% and 15.2% of the videos, respectively. The remainder of the websites include Instagram and Facebook as well as the websites of various global and local news organizations. For each of these videos, the URL of the video, title of the post, description of the post, and the date of publication of the video are presented as separate attributes in the dataset. After developing this dataset, sentiment analysis (using VADER), subjectivity analysis (using TextBlob), and fine-grain sentiment analysis (using DistilRoBERTa-base) of the video titles and video descriptions were performed. This included classifying each video title and video description into (i) one of the sentiment classes i.e. positive, negative, or neutral, (ii) one of the subjectivity classes i.e. highly opinionated, neutral opinionated, or least opinionated, and (iii) one of the fine-grain sentiment classes i.e. fear, surprise, joy, sadness, anger, disgust, or neutral. These results are presented as separate attributes in the dataset for the training and testing of machine learning algorithms for performing sentiment analysis or subjectivity analysis in this field as well as for other applications. Finally, this paper also presents a list of open research questions that may be investigated using this dataset.

## 1. Introduction

Measles is a highly transmissible viral illness caused by a single-stranded and enveloped RNA virus [1]. Despite the availability of an effective measles vaccine for more than 40 years, annually there are approximately 20 million cases of measles on a global scale and measles continues to be among the leading causes of death in young children [2,3]. The risk of measles has been significantly increased by the COVID-19 pandemic [4,5]. Furthermore, due to the impact of COVID-19 on the healthcare sector, from 2020 to 2022, more than 61 million doses of measles vaccines were missed or deferred on a global scale [6]. As a result, since the beginning of 2024, multiple countries have been experiencing outbreaks of measles. The countries include – Kazakhstan (21,740 cases), Azerbaijan (13,720 cases), Yemen (13,676 cases), India (13,220 cases), Iraq (11,595 cases), Ethiopia (9,042 cases), Kyrgyzstan (7,601 cases), Russian Federation (7,594 cases), Pakistan (5,812 cases), and Indonesia (5,648 cases). To add to this, the number of cases of measles in the United States since the beginning of 2024 has already exceeded the number of cases of measles recorded in the United States in 2023 [6].

Among the various types of web services and applications, online videos are currently "dominating the internet" [7]. On average, an individual watches 17 hours of videos on the internet per week [8] as online videos serve as a rich and seamless resource of information related to various topics including recent issues, global challenges, pandemics, virus outbreaks, emerging technologies, and trending matters [9]. In the last few years, social media platforms such as YouTube and TikTok have become popular amongst all groups as such platforms provide a seamless way for users to create and disseminate information in the form of videos [10,11]. On a global scale, YouTube is the second most frequented website on the internet after google.com. It is accessible in 100 nations and 80 languages, with users collectively streaming approximately 5 billion videos daily [12,13]. In terms of worldwide traffic on YouTube, the United States takes the lead with 11.67 billion, followed by South Korea (8.25 billion), India (4.2 billion), Brazil (3.59 billion), Germany (3.49 billion), and other countries [14]. In addition to this, more than 122 million individuals engage with YouTube daily, constituting roughly a quarter of global internet activity [15]. On a global scale, TikTok ranks 5[th] in the list of most popular social media platforms [16]. At present the number of active users on TikTok is 1.7 billion and this number is projected to increase to 2.25 billion by 2027 [17]. In 2024, TikTok has been the 3[rd] most downloaded mobile application on a global scale [18], and on average, each user spends 58 minutes and 24 seconds on TikTok on a daily basis [19]. In terms of the worldwide traffic on TikTok, the United States takes the lead with 148 million, which is followed by Indonesia (126.8 million), Brazil (98.6 million), Mexico (74.2 million), Vietnam (67.7 million), Russia (58.6 million), Pakistan (54.4 million), Philippines (49.1 million), Thailand (44.4 million), Turkey (37.7 million), and other countries [20].

During virus outbreaks of the recent past, social media platforms such as YouTube and TikTok served as crucial sources for the global population to stay informed and updated related to those virus outbreaks [21, 22]. Video datasets serve as valuable data resources for the investigation of diverse research questions related to creating, viewing, reacting, and disseminating video-based content on the internet. As a result in the last few years, researchers from different disciplines have focused on the development of datasets of videos published on YouTube, TikTok, and similar websites. The ongoing outbreak of measles which has been declared a public health emergency [23], epidemic [24], and a national incident [25] in different parts of the world has resulted in a concern about public health on a global scale. So, in the last few months, researchers from different disciplines have investigated the same as well as studied prior outbreaks of measles for insight related to the current outbreak. However, no prior work in this field has focused on the development of a dataset of videos about the ongoing outbreak of measles published on YouTube, TikTok, and other websites on the internet. To add to this, there are other research gaps that still exist in this field (discussed in Section 2). Addressing these gaps with an aim to contribute to the advancement of research in this field serves as the main motivation for this work. The rest of this paper is structured as follows. Section 2 presents a review of recent works in this field and discusses the research gaps that exist. Section 3 discusses the methodology that was followed for the development of this dataset. The results are presented in Section 4 which also includes a list of open research questions that may be investigated using this dataset. The conclusion is presented in Section 5 where the scientific contributions of this work are summarized and the scope for future work in this field is outlined.

## 2. Literature Review

Real et al. [26] developed a dataset that contains the URLs of YouTube videos for object detection. This dataset contains about 380,000 videos and the duration of each of these videos is approximately 19 seconds. Loh et al. [27] developed a dataset of YouTube videos for modeling internet traffic and streaming analysis. The dataset comprises 80 network scenarios, encompassing 171 distinct bandwidth settings. These settings were tested in a total of 5,181 tests with limited bandwidth, 1,939 runs with emulated 3G/4G traces, and 4,022 runs with pre-defined bandwidth variations. The work of Xu et al. [28] involved the development of a YouTube dataset for sequence-to-sequence video object segmentation. The dataset contains 3252 video clips and 78 types of common objects and human activities. Similar video datasets for object segmentation were developed by Li et al. [29], Jain et al. [30], Ochs et al. [31], Perazzi et al. [32], and Pont-Tuset et al. [33]. The number of videos these datasets contain is 14, 96, 59, 50, and 90, respectively. Lall et al. [34] collected watch history data of 243 YouTube users over a period of 1.5 years and developed a dataset. Their dataset contains a total of 1.8 million YouTube videos. Le et al. [35] developed a dataset of YouTube videos that contains 23,738 videos in four categories: comedy, travel and events, education, science and technology. Their dataset contains YouTube videos published over 12 years from 72 channels.

Qian et al. [36] developed a dataset of 283,582 TikTok videos for human activity recognition. The videos from this dataset represent 386 different hashtags related to human behavior. The work of Ng et al. [37] involved preparing a dataset of about 7000 videos from TikTok. The authors specifically focused on collecting videos where TikTok users demonstrated their participation or completion of trending challenges on TikTok. Basch et al. [38] collected 100 videos from TikTok containing #climatechange. 73 videos from this collection focused on at least one aspect of climate change. Fiallos et al. [39] developed a dataset of 1495 TikTok videos to understand the categories of knowledge and learning opportunities from TikTok. The dataset contains videos with different hashtags out of which #learnontiktok represents the primary hashtag for knowledge and learning opportunities. Shutsko et al. [40] developed a dataset of 1000 TikTok videos to analyze the trends of popularity of different subject matters on TikTok. The work of Abdaljaleel et al. [41] involved the assessment of information about the measles vaccine on social media platforms including YouTube and TikTok. The analysis of videos from these platforms showed that a majority of the videos (61.8%) were created by lay individuals and not medical professionals, healthcare providers, or journalists. Hussain et al. [42] performed an analysis of YouTube videos regarding measles. The findings showed that about 32% opposed vaccination against measles. Yiannakoulias et al. [43] analyzed content about measles vaccines as disseminated in YouTube videos. The findings from their analysis of 134 YouTube videos showed that 48.51% of the videos were in favor of getting vaccinated for measles, 19.40% of the videos were against getting vaccinated for measles, and 32.09% of the videos didn't communicate an opinion for or against getting vaccinated for measles.

To summarize, even though multiple works exist related to the development of datasets of YouTube videos, datasets of TikTok videos, and investigation of prior outbreaks of measles, there are multiple research gaps that exist in these areas of research. These research gaps are outlined as follows:

- No prior work in this field has focused on the development of a dataset of videos about the ongoing outbreak of measles published on YouTube, TikTok, and other websites on the internet.
- None of the video datasets in this field have attributes that assign an overall sentiment of positive, negative, or neutral to the video descriptions or video titles. To add to this, none of these datasets have attributes that assign a label such as anger, disgust, fear, joy, neutral, sadness, or surprise, to the video descriptions or video titles after performing fine-grain sentiment analysis.
- No prior work related to the development of a dataset of videos has attributes that categorize the video descriptions or video titles into one of the subjectivity classes - highly opinionated, neutral opinionated, and least opinionated, based on the degree of opinion expressed in each video.
- No prior work has presented the results of performing overall sentiment analysis, fine-grain sentiment analysis, or subjectivity analysis of videos related to the ongoing outbreak of measles from YouTube, TikTok, and other websites on the internet.
- None of these works that focused on the development of video datasets present datasets that contain the data of videos from YouTube, TikTok, Instagram, and Facebook as well as some of the popular news organizations such as cbsnews.com, nbcnews.com, msn.com, dailytelegraph.com.au, apnews.com, cnn.com.

The work presented in this paper aims to address these research gaps. The step-by-step methodology that was followed for the completion of this research work is discussed in Section 3.

## 3. Methodology

Figure 1 presents an overview of the methodology that was followed in this research work that resulted in the development of this dataset that contains the data of 4011 videos about the ongoing outbreak of measles published on 264 websites on the internet between January 1, 2024, and May 31, 2024, available at https://dx.doi.org/10.21227/40s8-xf63. These websites primarily include YouTube and TikTok, which account for 48.6% and 15.2% of the videos, respectively. The remainder of the websites includes Instagram and Facebook, as well as the websites of various global and local news organizations such as cbsnews.com, nbcnews.com, msn.com, dailytelegraph.com.au, apnews.com, cnn.com, etc.

For collecting data from YouTube, the YouTube API was used [44]. For the rest of the websites, the data was collected manually by the co-authors of this paper by using a keyword search on Google followed by visiting each of these websites. The keywords that were used for collecting the data included "measles" and "MMR vaccine". More specifically, if the title or description of a video contained either of these keywords that video was included in the development of the first version of the dataset. During the data collection process for the development of the first version of the dataset, for each video about measles, the URL of the video, the title of the post, the description of the post, and the date of publication of the video were collected. For websites such as tiktok.com, instagram.com, and a few news sources, as a separate video title and video description are not published, the value of the video title was used as the value of the video description. As this research work specifically focuses on the 2024 outbreak of measles, the time range for data collection was set as January 1, 2024, and May 31, 2024 (the most recent date at the time of submission of the camera-ready version of this paper to HCII 2024) and the data of videos published before January 1, 2024, were removed from the dataset. Thereafter, data preprocessing of the video titles and video descriptions was performed by writing a program in Python 3.11.5 installed on a computer with a Microsoft Windows 10 Pro operating system. The data preprocessing included (a) removal of characters that were not alphabets, (b) removal of URLs, (c) removal of hashtags, (d) removal of user mentions, (e) detection of English words using tokenization, (f) stemming, g) removal of stop words, and (h) removal of numbers.

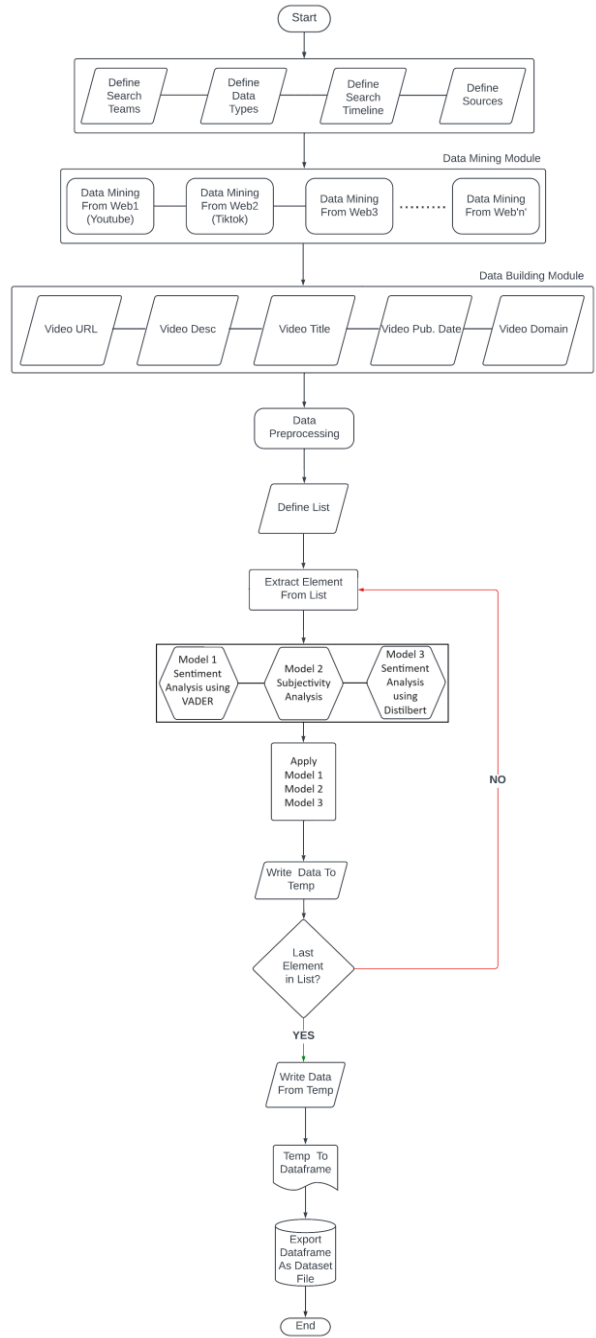

**Figure 1**. A flowchart that represents an overview of the methodology that was followed for the development of this dataset.

Finally, edge cases were also removed from the dataset by manual review of the video titles and video descriptions. This manual review was performed by the co-authors of this paper. In this context, we define edge cases as video titles or video descriptions that met our search criteria but were not related to the ongoing outbreak of measles (for example [45]). Thereafter, the preprocessed versions of the video titles and video descriptions were analyzed using VADER for Sentiment Analysis [46], TextBlob for Subjectivity Analysis [47], and DistilRoBERTa-base for fine-grain sentiment analysis [48].

VADER (Valence Aware Dictionary and sEntiment Reasoner) is a lexicon and rule-based sentiment analysis tool that is specifically attuned to sentiments expressed in social media [46]. VADER can analyze a given text and categorize it as either positive, negative, or neutral. In addition, it can identify the compound sentiment score and the magnitude of sentiment represented in a given text, ranging from 0 to +4 for positive sentiment and 0 to -4 for negative sentiment. There are multiple factors as a result of which VADER was used for sentiment analysis in this work even though several other approaches for sentiment analysis exist. First, studies have shown that VADER demonstrates outstanding efficiency with respect to both precision and efficacy [49-51]. Second, VADER effectively addresses the limitations faced by several other sentiment analysis approaches [52-55]. Finally, VADER has attracted the attention of researchers from different disciplines for solving research problems that focused on performing sentiment analysis of conversations on the internet related to recent virus outbreaks [56-60]. TextBlob is a lexicon-based analyzer that uses a set of predefined rules to perform sentiment analysis and subjectivity analysis. The sentiment score lies between −1 to 1, where −1 identifies the most negative words such as 'disgusting', 'awful', and 'pathetic', and 1 identifies the most positive words like 'excellent', and 'best'. The subjectivity score lies between 0 and 1. It represents the degree of personal opinion, if a sentence has high subjectivity i.e., close to 1, it means that the text contains more personal opinion than factual information. For fine-grain sentiment analysis, the specific model that was used was DistilRoBERTa-base [48]. This model can analyze a text and categorize it into one of the fine-grain sentiment classes - anger, disgust, fear, joy, neutral, sadness, or surprise. This model is a fine-tuned checkpoint of DistilRoBERTa-base and has been used in multiple prior works in this field that involved performing fine-grain sentiment analysis [61-63]. The results of applying VADER to the descriptions and titles of these videos were compiled and added as two new attributes – "VADER_Description" and "VADER_Title" to the dataset. These two attributes present the classification of video descriptions and video titles as positive, negative, or neutral using VADER. The results of subjectivity analysis were also compiled and added as two new attributes - "Subjectivity_Description" and "Subjectivity_Title", where the video descriptions and video titles are classified as Highly Opinionated, Neutral Opinionated, or Least Opinionated. These subjectivity or opinion classes based on the output from TextBlob were defined based on multiple prior works in this field where TextBlob was used for performing subjectivity analysis (for example: [64,65]).

Finally, the results of applying DistilRoBERTa-base to the descriptions and titles of these videos to perform fine-grain sentiment analysis were also compiled and added as two different attributes – "FineGrainSentiment_Description" and "FineGrainSentiment_Title", where the video descriptions and video titles are classified as anger, disgust, fear, joy, neutral, sadness, or surprise. These results are discussed in detail in Section 4, which also presents the results of data analysis and a list of open research questions that may be investigated using this dataset.

## 4. Results and Discussions

This section presents the results of this research work. The dataset that was developed is available at https://dx.doi.org/10.21227/40s8-xf63. This dataset is compliant with the FAIR (Findability, Accessibility, Interoperability, and Reusability) principles for scientific data management [66]. The dataset is findable, as it has a unique and permanent DOI, which has been assigned by IEEE Dataport. The dataset can be accessed online by any individual on the internet by directly visiting the DOI of the dataset. It is interoperable due to the use of a .csv file that can be downloaded, read, and analyzed across different operating systems and applications. The dataset is reusable as the video-related information, such as the URLs of the videos, titles of the posts, descriptions of the posts, the dates of publication of the videos, overall sentiment classes, fine-grain sentiment classes, and subjectivity classes from the dataset file can be used for free for the development of any types of programs or algorithms any number of times without any requirement to purchase any subscription or credits per use.

The results of the data analysis are shown in Figures 2-7. The results of sentiment analysis using VADER in Figures 2 and 3 show that for the video titles, 62.78% were neutral, 20.04% were positive, and 17.18% were negative and for the video descriptions 40.46% were neutral, 39.42% were positive, and 20.12% were negative. The results of subjectivity analysis using TextBlob from Figures 4 and 5 show that for the video titles, the distribution of the classes highly opinionated, neutral opinionated, and least opinionated were 5.93%, 17.85%, and 76.22%, respectively, and for the video descriptions, the distribution of these opinion classes were 10.07%, 27.25%, and 62.68%, respectively. The results of sentiment analysis using DistilRoBERTa-base showed that for the video descriptions, the distribution of fine-grain sentiment classes of fear, surprise, joy, sadness, anger, disgust, and neutral was 26.18%, 2.17%, 3.44%, 8.63%, 2.34%, 0.42%, and 56.82%, respectively and for the video titles the distribution of these fine-grain sentiment classes was 18.37%, 2.22%, 1.20%, 6.31%, 2.12%, 0.82%, and 68.96%, respectively. In this context, the authors would like to clarify that the sentiment class, subjectivity class, and fine grain sentiment class assigned to each video title and video description in this dataset, are presented in *as-is* form after obtaining the same from the outputs of VADER, TextBlob, and DistilRoBERTa-base, respectively. These outputs as well as the video data present in this dataset do not represent or reflect the views, opinions, beliefs, or political stances of the authors of this paper.

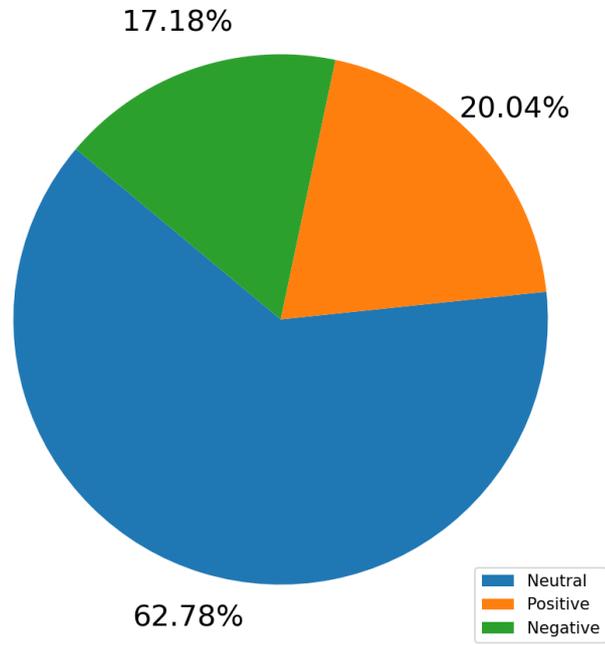

**Figure 2**: Results of Sentiment Analysis of the Video Titles using VADER

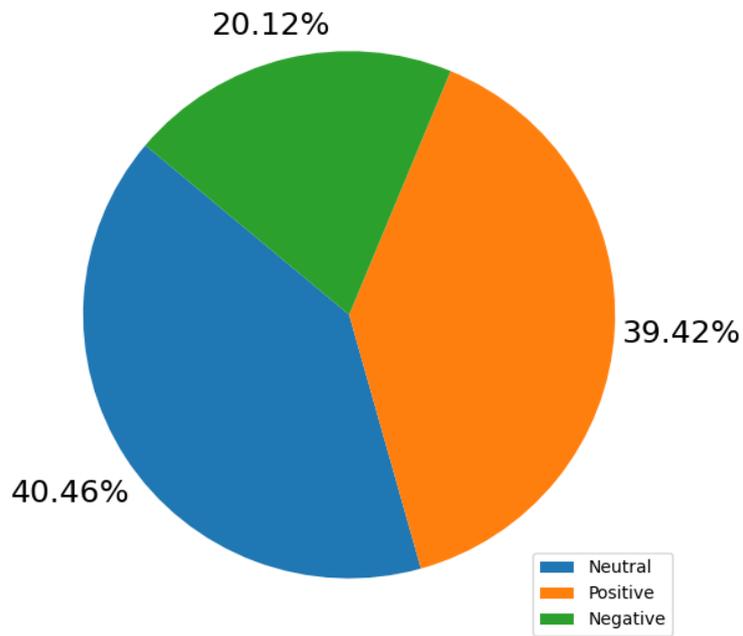

**Figure 3**: Results of Sentiment Analysis of the Video Descriptions using VADER

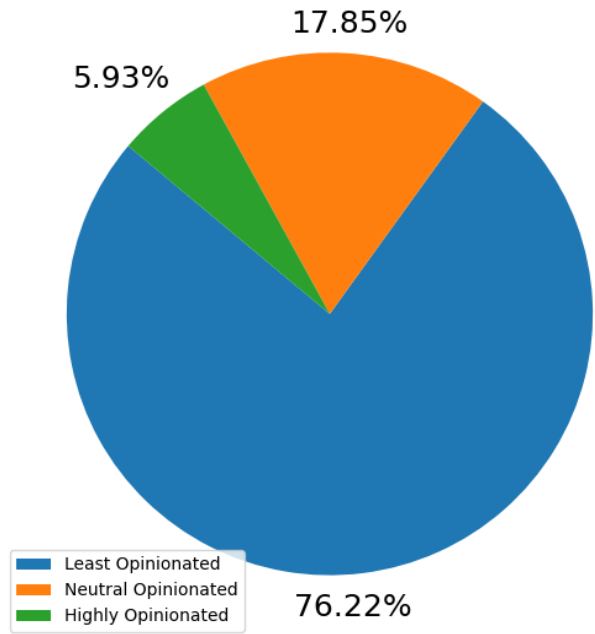

**Figure 4**: Results of Subjectivity Analysis of the Video Titles using TextBlob

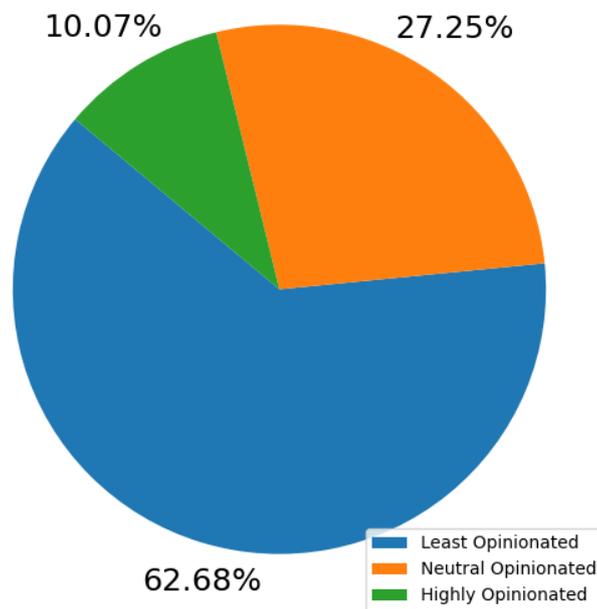

**Figure 5**: Results of Subjectivity Analysis of the Video Descriptions using TextBlob

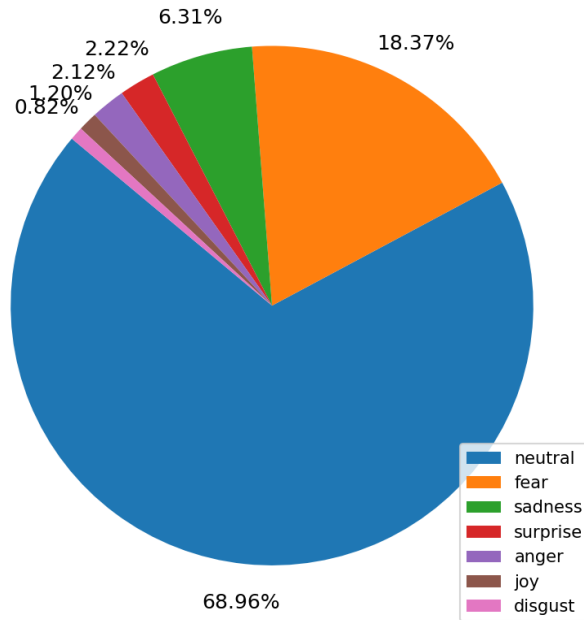

**Figure 6**: Results of Fine-Grain Sentiment Analysis of the Video Titles using DistilRoBERTa-base

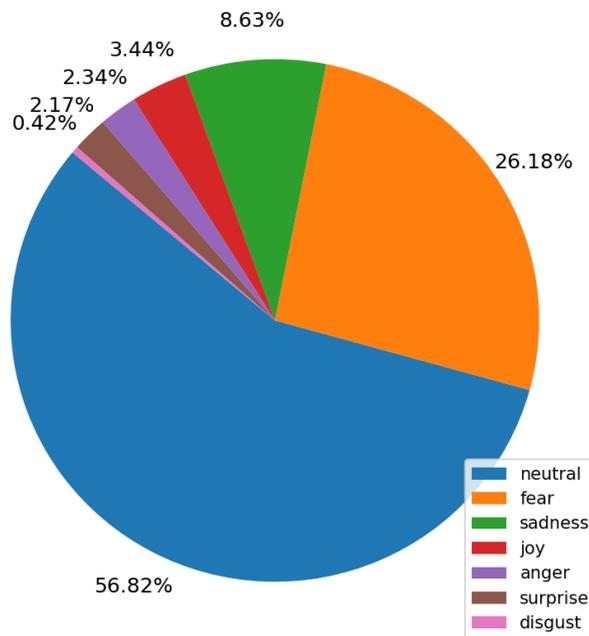

**Figure 7**: Results of Fine-Grain Sentiment of the Video Descriptions using DistilRoBERTa-base

YouTube and TikTok, being among the top five globally popular social media platforms are popular for video-based content creation and dissemination on a wide range of topics. Prior works have shown that the nature and intensity of sentiments on YouTube and TikTok vary from topic to topic [67-71]. For instance, the work of Shevtsov et al. [72] showed that in the context of the 2020 presidential elections in the United States, the sentiment towards the presidential candidates was predominantly negative. In [73], the authors concluded that in terms of conspiracy theories related to COVID-19 on YouTube, the distribution of sentiment was 46.9% positive, 31.0% neutral, and 22.1% negative. In [74], the authors showed that the sentiment towards vaccinations on YouTube was primarily negative (52%) in 2017. However, this changed to primarily positive (54%) in 2018. The work of Rachmawati et al. [75] showed that TikTok videos containing #samasamabelajar were primarily neutral (57.06%). In this paper, the findings of sentiment analysis show that the percentage of videos with neutral sentiment is higher than the percentage of videos with other sentiments. There may be multiple reasons that support this finding. First, the videos of this dataset have been published on YouTube, TikTok, Instagram, Facebook, and other websites such as cbsnews.com, nbcnews.com, msn.com, dailytelegraph.com.au, apnews.com, cnn.com, etc. since the beginning of this year. Many of these websites are news channels that are likely to present the facts as compared to presenting their opinions (positive or negative) related to the ongoing measles outbreak. For a video that presents only factual information, the assigned label from sentiment analysis using VADER and DistilRoBERTa-base would be neutral. Second, the ongoing outbreak of measles, similar to the virus outbreaks of the recent past, isn't a topic for which the global population is expected to have negative opinions (unlike the 2020 Presential Elections in the United States as per the findings of Shevtsov et al. [72]). As a result, the majority of the videos in this dataset are not associated with a negative sentiment.

A list of open research questions is presented next, which may be investigated using this dataset:

- o Performing topic modeling of the video descriptions to identify the themes or focus areas related to the creation and dissemination of videos about the ongoing outbreak of measles.
- o Performing aspect-based sentiment analysis of these videos to investigate the specific topics or focus areas regarding the ongoing measles outbreak associated with positive or negative sentiments or one of the fine-grain sentiment classes of fear, surprise, joy, sadness, anger, or disgust.
- o Performing a case study of different supervised learning models in machine learning to determine the optimal model for identification of the overall sentiment, the fine-grain sentiment, or the subjectivity expressed in these videos.
- o Detecting sarcasm expressed in video descriptions or video titles to identify the trends of sarcasm (using the dates of publications of the videos that are present in the dataset) about the ongoing outbreak of measles on social media platforms such as YouTube, TikTok, Instagram, and Facebook and analyzing any similarities or differences in those trends.

- Investigation of any correlations between the length of video titles (or descriptions) and the overall sentiment, fine-grain sentiment, or subjectivity in the videos.
- Analyzing the usage of hashtags in video descriptions related to the ongoing outbreak of measles on YouTube and TikTok to identify the popular hashtags associated with positive, negative, and neutral videos.
- Detecting distinct users (from the video URLs that are present in the dataset) who published these videos on TikTok and YouTube. Thereafter, identifying the types of users (for example, medical professionals, healthcare organizations, etc.) who posted the majority of positive, negative, and neutral videos.
- Performing Content Value Analysis and analysis of the credibility of information in these videos to rank the video sources from highest credibility to lowest credibility regarding information about the ongoing measles outbreak on social media platforms such as YouTube, TikTok, Instagram, and Facebook.
- Develop a binary classifier to categorize the video sources as a news source or not a news source. Thereafter, identifying the role of news sources in the dissemination of information about the ongoing outbreak of measles.
- Detecting misinformation expressed by the different video sources to analyze the degree of misinformation dissemination per video source related to the ongoing outbreak of measles.
- Detecting fake news regarding measles expressed by the different video sources to analyze the degree of misinformation dissemination per video source.
- Identification of conspiracy theories expressed in videos about the ongoing measles outbreak to infer which platform(s) has been playing a greater role in the creation and dissemination of conspiracy theories.
- Detecting satire related to measles expressed in video descriptions and the trends of the satire on social media platforms such as YouTube, TikTok, Instagram, and Facebook on a weekly or biweekly basis (using the date of publication of these videos which is present in the dataset).
- Identification of hate speech or abusive language in videos about the ongoing outbreak of measles published on social media platforms such as YouTube, TikTok, Instagram, and Facebook to determine the trends of the same.
- Detection, identification, and ranking of trending topics for video publication on the internet related to the ongoing outbreak of measles.
- Detection of communities on social media platforms such as YouTube, TikTok, Instagram, and Facebook that support or do not support each other regarding the ongoing outbreak of measles based on the analysis of reaction videos and related characteristics.

This dataset and the open research questions presented in this paper are expected to advance research and development in this field. Furthermore, the findings of this paper are also expected to contribute to the development of video recommendations related to the ongoing outbreak of measles. The methodology used by modern-day video recommendation systems is either collaborative filtering or content-based filtering [76]. Collaborative filtering algorithms evaluate components of user behavior related to watching videos such as ratings, likes, dislikes, watch time, favorites, etc. to create a profile of each user as per their interests. Then, the algorithm pairs the user with other users with similar behavior. Thereafter, it analyzes similar behaviors to develop video recommendations for similar users [77]. However, collaborative filtering-based video recommendations have a "cold start" problem as many users on the internet do not like, dislike, or rate videos. Such systems also require a considerable amount of data for the identification of similar users [78,79]. Therefore, content-based filtering approaches for video recommendations have been gaining popularity in the recent past [80]. Content-based filtering approaches take into account multiple characteristics of videos for recommending videos to users. The concept of content-based video recommendations is used by multiple video streaming platforms [80]. As this paper presents the findings of sentiment analysis, fine-grain sentiment analysis, and subjectivity analysis and the assignment of sentiment, fine-grain sentiment, and subjectivity classes to each video in this dataset, this work is also expected to contribute towards the development of content-based video recommendation systems related to the ongoing measles outbreak.

The work presented in this paper has a few limitations. First, VADER and DistilRoBERTa-base were used for performing sentiment analysis and fine-grain sentiment analysis. To add to this, TextBlob was used for performing subjectivity analysis. These three algorithms use unsupervised learning and have been widely used for sentiment analysis, fine-grain sentiment analysis, and subjectivity analysis in several prior works in this field [81-86]. However, none of these algorithms are 100% accurate. Second, other than YouTube, the video data (i.e. the URL of the video, title of the post, description of the post, and the date of publication of the video) from sources such as TikTok, Instagram, Facebook, and websites of global and local news channels, presented in this dataset was collected by the co-authors of this paper by manually visiting different websites (discussed in Section 3). As stated in prior works where manual labeling was used [87,88], manual labeling may be associated with minor human errors. Third, even though it is not stated in the description of the DistilRoBERTa-base model [48], we observed that this model is able to process up to 512 characters for performing fine-grain sentiment analysis. To address this limitation of the model, for video descriptions and video titles that exceeded its processing limit, we passed the first 512 characters to the model. Finally, the findings of sentiment analysis, fine-grain sentiment analysis, and subjectivity analysis as presented in this paper are based on the videos that are available in this dataset. The ongoing outbreak of measles continues to affect multiple geographic regions of the world. As a result, multiple videos related to this outbreak are getting published on the internet every day. Therefore, if sentiment analysis, fine-grain sentiment analysis, and subjectivity analysis of videos about the ongoing outbreak of measles are performed at any time in the near future, depending on the global reaction, views, opinions, and responses towards the outbreak at that time, the results may vary as compared to the results presented in this paper.

## 5. Conclusion

Measles is a highly contagious viral illness produced by a single-stranded RNA virus and prior works in this field have shown a substantial rise in the susceptibility to measles as a direct consequence of the COVID-19 pandemic. Since the beginning of 2024, several countries have been experiencing an outbreak of measles. Online videos are currently exerting a dominant influence on the internet. Social media platforms such as YouTube and TikTok have gained widespread popularity across many demographics on a global scale due to their ability to facilitate the easy creation and sharing of videos. During virus outbreaks of the recent past, social media platforms such as YouTube and TikTok played a vital role in keeping the worldwide public informed and up to date on the outbreaks. Video datasets serve as valuable data resources for the investigation of diverse research questions related to the creation and dissemination of video-based content on the Internet. As a result in the last few years, researchers from different disciplines have focused on the development of datasets of videos published on YouTube, TikTok, and similar websites. However, no prior work in this field has focused on the development of a dataset of videos about the ongoing outbreak of measles published on YouTube, TikTok, and other websites on the internet. To add to this, there are other research gaps that still exist in this field. The work of this paper aims to address these research gaps and presents a dataset that contains the data of 4011 videos about the ongoing outbreak of measles uploaded on 264 websites on the internet between January 1, 2024, and May 31, 2024, available at https://dx.doi.org/10.21227/40s8-xf63. These websites primarily include YouTube and TikTok, which account for 48.6% and 15.2% of the videos, respectively. The remainder of the websites include Instagram and Facebook as well as the websites of multiple global and local news organizations such as cbsnews.com, nbcnews.com, msn.com, dailytelegraph.com.au, apnews.com, cnn.com, etc. For each of these videos, the URL of the video, title of the post, description of the post, and the date of publication of the video are presented as separate attributes in the dataset. The work of this paper also included performing sentiment analysis (using VADER), subjectivity analysis (using TextBlob), and fine-grain sentiment analysis (using DistilRoBERTa-base) of the video titles and video descriptions. These results are presented as separate attributes in the dataset. The dataset complies with the FAIR principles of scientific data management. The paper also presents a list of open research questions that may be investigated using this dataset. As per the best knowledge of the authors, no similar work has been done in this field thus far. Future work in this area would include extending the dataset as well as investigating the presented research questions and research directions.

**Disclosure of Interests**. The authors have no competing interests to declare that are relevant to the content of this article.

18